\documentclass[useAMS,usenatbib]{mnras}
\usepackage[english]{babel}
\usepackage[fleqn]{amsmath}
\usepackage{amsfonts}
\usepackage{txfonts}
\usepackage{graphicx}
\usepackage{amssymb}
\usepackage[T1]{fontenc}
\usepackage{ae,aecompl}
\usepackage{fancyhdr}
\usepackage{multicol}
\usepackage{layout}
\usepackage{graphicx}
\usepackage{multirow}
\usepackage{color}
\usepackage{multirow}
\usepackage{times}
\usepackage[outdir=./fig/]{epstopdf}

\graphicspath{{./fig/}}

\newif\ifAMStwofonts
\AMStwofontstrue

\title[Spherical collapse with tidal shear]{Effects of tidal gravitational fields in clustering dark energy models}
\author[Pace et~al.]{Francesco Pace$^{1}$\thanks{e-mail:francesco.pace@manchester.ac.uk}, Robert Reischke$^{2}$, 
Sven Meyer$^3$ and Bj\"orn Malte Sch\"afer$^{2}$\\
$^{1}$ Jodrell Bank Centre for Astrophysics, School of Physics and Astronomy, The University of Manchester, Manchester, 
M13 9PL, United Kingdom\\
$^{2}$ Zentrum f{\"u}r Astronomie der Universit{\"a}t Heidelberg, Astronomisches Recheninstitut, Philosophenweg 12, 
69120 Heidelberg, Germany\\
$^{3}$Zentrum f{\"u}r Astronomie der Universit{\"a}t Heidelberg, Institut f{\"u}r theoretische Astrophysik, 
Philosophenweg 12, D-69120, Heidelberg, Germany}

\date{Accepted ?, Received ?; in original form \today}

\pagerange{\pageref{firstpage}--\pageref{lastpage}} \pubyear{2016}

\begin{document}

\label{firstpage}

\maketitle

\begin{abstract}
We extend a previous work by Reischke et al., 2016 by studying the effects of tidal shear on clustering dark energy 
models within the framework of the extended spherical collapse model and using the Zel'dovich approximation. As in 
previous works on clustering dark energy, we assumed a vanishing effective sound speed describing the perturbations in 
dark energy models. To be self-consistent, our treatment is valid only on linear scales since we do not intend to 
introduce any heuristic models. This approach makes the linear overdensity $\delta_{\rm c}$ mass dependent and 
similarly to the case of smooth dark energy, its effects are predominant at small masses and redshifts. 
Tidal shear has effects of the order of percent or less, regardless of the model and preserves a well known feature of 
clustering dark energy: When dark energy perturbations are included, the models resemble better the $\Lambda$CDM 
evolution of perturbations. We also showed that effects on the comoving number density of halos are small and 
qualitatively and quantitatively in agreement with what previously found for smooth dark energy models.
\end{abstract}


\section{Introduction}
A wealth of data accumulated over the last two decades \citep{Riess1998,Perlmutter1999,Cole2005,Planck2016_XIII} 
strongly hints to the conclusion that the expansion of the Universe is currently accelerated and its origin represents 
one of the largest challenges and open questions in both physics and cosmology. The simplest explanation is the 
cosmological constant $\Lambda$, which, so far, is very consistent with the data 
\citep{Planck2016_XIII,Planck2016_XIV}. Despite this agreement, a conclusive answer is still lacking and to refuse or 
confirm the standard cosmological model, dubbed $\Lambda$CDM model, where $\Lambda$ is the cosmological constant 
responsible for the accelerated expansion and CDM represents the cold dark matter component responsible for the 
majority of the gravitating mass in the Universe, it is necessary to investigate other cosmological models.

Staying within the framework of General Relativity, the simplest extension of the basic $\Lambda$CDM model is the 
introduction of a fluid with a time-varying equation of state $w<-1/3$ which is then called dark energy. Many models 
have been proposed and two different approaches can be considered. The simplest one is to propose a phenomenological 
equation of state $w(a)$ which will depend on some parameters that need to be fitted via observations 
\citep{Chevallier2001,Linder2003,Hannestad2004,Jassal2005,Barboza2008}. This approach is purely phenomenological and 
there is no model behind to explain why the proposed equation of state must have that particular functional form. The 
second one is more theoretical and it is based on the identification of the dark energy component with a scalar field, 
since in this framework a time-dependent equation of state arises naturally. The equation of state is defined as 
$w_{\phi}=P_{\phi}/\rho_{\phi}$ where $P_{\phi}$ and $\rho_{\phi}$ are the pressure and the energy density of the 
scalar field. A similar definition holds for the fluid formalism as well, upon the identification of the pressure and 
the energy density of the scalar field with the corresponding quantity for the dark energy fluid. In this setup, it is 
necessary to specify a self-interacting potential and by solving the corresponding Klein-Gordon equation (which 
describes the time evolution of the scalar field) it is possible to recover the corresponding equation of state 
\citep[see e.g.][for a review]{Copeland2006,Tsujikawa2010}. 
In this case the parameters to be fitted are those of the scalar field potential. 
For a recent discussion about the background properties of the most studied scalar field models we refer to 
\cite{Battye2016} and references therein.

While the cosmological constant only affects the background evolution described by the Hubble function $H(a)$, dark 
energy models possess fluctuations also on sub-horizon scales if $w\ne -1$, despite being small and practically 
negligible with respect to matter perturbations. Dark energy perturbations will therefore affect also the evolution of 
matter perturbations and the determination of their effects is one of the goals of future surveys. 
Perturbations in the dark energy sector are encoded by the effective sound speed $c_{\rm eff}^2$, defined as 
$c_{\rm eff}^2\equiv \delta P/\delta\rho$ where $\delta P$ and $\delta\rho$ represent the pressure and density 
perturbation of the fluid. Note that for matter the same definition will yield $c_{\rm eff}^2=0$, therefore we can 
identify a fluid as fully clustering when its effective sound speed is null. To infer its correct functional form, once 
again we need a theoretical background model. For quintessence models and phantom models, $c_{\rm eff}^2=1$ always, 
therefore the perturbations are negligible and the dark energy component is homogeneous. For {\it k}-essence models, 
the effective sound speed can be close to zero and significantly different from unity, but one has to specify a 
functional form for both the potential and the kinetic term.

Dark energy perturbations will affect not only the linear dynamics of perturbations via the growth factor, but also 
their non-linear evolution and this will reflect on the mass function which is of fundamental importance for cluster 
counts \citep{Majumdar2004,Diego2004,Fang2007,Abramo2009a,Angrick2009} and weak lensing peak counts 
\citep{Maturi2010,Maturi2011,Reischke2016}. The inclusion of dark energy perturbations into the non-linear dynamics 
leading to the evaluation of the mass function is usually done via the spherical collapse model 
\citep{Fillmore1984,Bertschinger1985,Ryden1987,AvilaReese1998,Mota2004,Abramo2007}, according to which perturbations 
describe a spherically symmetric and non-rotating object which, being overdense, decouples from the background 
acceleration, reaches a maximum radius (turn around) and eventually collapses to a point. In reality a null size is 
never reached since a virial equilibrium is created. The mass function is sensitive to the normalization of the matter 
power spectrum $\sigma_8$ and the matter density parameter $\Omega_{\rm m}$ \citep{Angrick2015}, to the dark energy 
equation of state and its evolution 
\citep{Holder2001,Haiman2001,Weller2002,Majumdar2003,Pace2012} and in general to the background cosmological model 
\citep{Pace2014}. The mass function is also a key ingredient in the halo model formalism for both dark matter halos and 
halos detected via the Sunyaev-Zel'dovich effect.

It is therefore important to study how the mass function gets modified when additional features are added to the 
standard picture: Previous works, based on a phenomenological approach, showed that deviation from isotropic shear and 
non-zero rotation of the perturbed object play an important role into the non-linear evolution of matter perturbations 
and this has a tremendous effect on the resulting cluster number counts 
\citep{DelPopolo2013a,DelPopolo2013b,DelPopolo2013c,Pace2014b}. 
More recently, \cite{Reischke2016a}, using the Zel'dovich approximation to theoretically model tidal shear interactions 
in dark energy models, showed that when tidal shear is taken into account, a $1-\sigma$ bias in the determination of 
the cosmological parameters can arise. It is therefore important to quantify this effect also for clustering dark 
energy model, since a precise determination of the equation of state parameter and eventually of its effective sound 
speed is the goal of future surveys such as Euclid\footnote{\url{http://www.euclid-ec.org/}} 
\citep{Laureijs2011,Amendola2013}, LSST\footnote{\url{http://www.lsst.org}} \citep{LSST2012} and 
SKA\footnote{\url{https://www.skatelescope.org/}} \citep{Bull2015,Camera2015,Santos2015}.

The structure of the paper is as follows. In Sect.~\ref{sect:scm} we review the extension of the simple spherical 
collapse model to include dark energy perturbations and tidal shear effects. In Sect.~\ref{sect:Tidal} we summarize the 
statistical procedure we used to sample the tidal shear values required to study the non-linear evolution of 
perturbations. Sect.~\ref{sect:DEmodels} is devoted to the study of the effect of tidal shear on selected dark energy 
models whose results will be the starting point for the discussion in Sect.~\ref{sect:mf}, where we discuss the effects 
on the mass function and on the cluster counts. Finally in Sect.~\ref{sect:discussion} we discuss and summarize our 
results.

\section{Spherical Collapse Model}\label{sect:scm}
The formalism of the spherical collapse model allows the study of the evolution of the perturbations of cosmological 
fluids into the non-linear regime and as such it is an important tool in cosmology and in the study of structure 
formation. This model, initially proposed for the Einstein-de Sitter (EdS) model \citep{Gunn1972}, has been studied and 
discussed by several authors \citep{Padmanabhan1993,Padmanabhan1996,Fosalba1998a,Mota2004,Nunes2006,Anselmi2011}. It 
assumes spherically symmetric, isolated and non-rotating overdense regions detaching from the background expansion, 
reaching a maximum size and then recollapsing again. Despite its simplicity, it has proven a powerful and accurate tool 
and it is at the basis of the mass function formalism, as we will discuss later.

The formalism of the spherical collapse model has been extended to include additional effects and make it more 
realistic. \cite{DelPopolo2002,Ohta2003,Ohta2004}, for example, extended it to deviations from sphericity allowing an 
ellipsoidal geometry. \cite{Abramo2007,Abramo2009a,Basilakos2009,Creminelli2010,Batista2013} applied the formalism to 
clustering dark energy models, while \cite{Basilakos2010} studied time varying vacuum cosmologies. 
A survey of dark energy models, including early dark energy models, is presented in 
\cite{Szydlowski2006,Jennings2010,Pace2010} and a detailed study of several functional forms for oscillating dark 
energy models is performed in \cite{Pace2012}. 
More recently, the spherical collapse model has been applied to scalar-tensor theories (non-minimally coupled models) 
\citep{Pace2014,NazariPooya2016}, to coupled dark energy models \citep{Wintergerst2010a,Tarrant2012} and neutrino 
cosmologies \citep{LoVerde2014}.

\cite{DelPopolo2013a} and \cite{DelPopolo2013b} included, for several dark energy models, the effects of shear and 
rotation and its extension to clustering dark energy models can be found in \cite{Pace2014b}. These last works, with 
the inclusion of the contribution from shear and rotation, are based on a purely phenomenological approach. 
A way out, based on the Zel'dovich approximation can be found in \cite{Reischke2016a}.

In this section we review the equations of the spherical collapse model for clustering dark energy, writing explicitly 
the contribution of the effective sound speed $c_{\rm eff}$ as done in \citet{Basse2011}, even if in the following, in 
agreement with other works on the subject, we will set it to zero, therefore considering the full-clustering case. For 
a full derivation we refer to \cite{Abramo2007,Abramo2009a}.

The effective sound speed $c_{\rm eff}$ is in general a function of scale and time and it represents a new degree of 
freedom describing dark energy perturbations. To fully evaluate it, it is necessary to have a model to work with. In 
the limit of small scales, as the ones we are considering here, it is possible to assume that it only depends on time 
and general expressions for scalar-tensor models within the Horndeski class are given in \cite{Gleyzes2013} and 
\cite{Gleyzes2014}. In addition, $c^2_{\rm eff}\geq 0$ to avoid instabilities and $c^2_{\rm eff}\leqslant 1$ for the 
propagation of perturbations to be sub-luminal. The limiting case $c^2_{\rm eff}=1$ ($c^2_{\rm eff}=0$) corresponds to 
a smooth (fully clustering) model and in this case dark energy perturbations are much smaller than (comparable to) the 
dark matter ones, as shown in \cite{Abramo2007} and \cite{Batista2013}.

A further reason leading to the choice of setting the effective sound speed to zero is given in \cite{Abramo2009b}: in 
this limiting case, the equations in the linear regime coincide for both the Newtonian and the general relativistic 
case.

Said this, it is nevertheless illuminating to understand what happens when the effective sound speed differs from the 
limiting cases. To do so, we refer to \cite{Basse2011} and \cite{Basse2012} (their Fig.~2): the authors showed that 
decreasing the value of the effective sound speed, the effect of dark energy perturbations increases. To understand 
this we can repeat the analysis of \cite{Batista2013}: the qualitative behaviour of dark energy perturbations is 
determined by two scales, the particle horizon
\begin{equation}
 \lambda_{\rm H}(a) = \int_{a_{\rm i}}^a\frac{da^{\prime}}{{a^{\prime}}^2H} \;,
\end{equation}
and the sound horizon
\begin{equation}
 \lambda_{\rm s}(a) = \int_{a_{\rm i}}^a\frac{c_{\rm eff}da^{\prime}}{{a^{\prime}}^2H} \;.
\end{equation}
If $\lambda<\lambda_{\rm s}$ perturbations oscillate with decreasing amplitude, while for $\lambda>\lambda_{\rm s}$ 
perturbations are effectively pressureless and will grow at the same pace of matter perturbations. 
We can further define a non-linear scale, typical of galaxy clusters, and set it to 
$\lambda_{\rm nl}\simeq 10h^{-1}$Mpc, corresponding to $k_{\rm nl}\simeq 0.63h$Mpc$^{-1}$. 
Setting $k_{\rm s}=k_{\rm nl}$ today and assuming a constant effective sound speed for simplicity, we find 
$c_{\rm nl}\approx 10^{-3}$. When $c_{\rm eff}\ll c_{\rm nl}$, $\lambda_{\rm s}\ll\lambda_{\rm nl}$, dark energy 
perturbations will be pressureless and we have effectively $c_{\rm eff}=0$. If instead $c_{\rm eff}\gg c_{\rm nl}$, 
dark energy perturbations are strongly suppressed by their pressure support.

We can therefore consider our analysis as the limiting case where the effects of perturbations in dark energy are at 
their maximum.

The perturbed continuity equations for matter (here with matter we consider both cold dark matter and baryons) and dark 
energy are
\begin{align}
 \delta_{\rm m}^{\prime}+\left(1+\delta_{\rm m}\right)\frac{\theta}{aH} = & \; 0\;,\label{eqn:DMp}\\
 \delta_{\rm de}^{\prime}+\frac{3}{a}\left(c_{\rm eff}^2-w_{\rm de}\right)\delta_{\rm de}+
 \left[1+w_{\rm de}+\left(1+c_{\rm eff}^2\right)\delta_{\rm de}\right]\frac{\theta}{aH} = &\; 0\;,\label{eqn:DEp}
\end{align}
where $w$ is the equation of state for the dark energy component, $\delta$ and $\theta$ the density contrast and the 
divergence of the peculiar comoving velocity of the fluid, respectively, and a prime represents the derivative with 
respect to the scale factor $a$.

Assuming that tidal field and rotation effects affect both fluids equally, Euler equations reads
\begin{equation}
 \theta^{\prime}+\frac{2}{a}\theta+\frac{\theta^2}{3aH}+
 \frac{\sigma^2-\omega^2}{aH}+
 \frac{3H}{2a}\left[\Omega_{\rm m}\delta_{\rm m}+\Omega_{\rm de}\delta_{\rm de}\right] = 0\;.\label{eqn:thetap}
\end{equation}
If this were not the case, we would have to consider a different evolution for each $\theta$. In this case, 
$\theta_{\rm m}\neq\theta_{\rm de}$ and the two Euler equations will be
\begin{align}
 \theta_{\rm m}^{\prime}+\frac{2}{a}\theta_{\rm m}+\frac{\theta_{\rm m}^2}{3aH}+
 \frac{\sigma_{\rm m}^2-\omega_{\rm m}^2}{aH}+
 \frac{3H}{2a}\left[\Omega_{\rm m}\delta_{\rm m}+\Omega_{\rm de}\delta_{\rm de}\right] = &\; 0\;,\label{eqn:thetaDMp}\\
 \theta_{\rm de}^{\prime}+\frac{2}{a}\theta_{\rm de}+\frac{\theta_{\rm DE}^2}{3aH}+
 \frac{3H}{2a}\left[\Omega_{\rm m}\delta_{\rm m}+\Omega_{\rm de}\delta_{\rm de}\right] = &\; 0\;.\label{eqn:thetaDEp}
\end{align}
In the previous equations, $\sigma$ and $\omega$ represent the contribution of the tidal field and of the vorticity, 
respectively. We will show how to evaluate $\sigma$ (or eventually $\sigma_{\rm m}$) in Sect.~\ref{sect:Tidal}, while 
in the rest of this work we will consider a vanishing net vorticity ($\omega=0$). We will consider both fluids affected 
in the same way by $\sigma$ and in Sect.~\ref{sect:DEmodels} we will outline the differences and consequences of the 
two assumptions.

Note that it is more convenient to work with dimensionless quantities, therefore we define $\tilde{\theta}=\theta/H$. 
Making the appropriate substitutions, the system of equations we need to solve is:
\begin{align}
 \delta_{\rm m}^{\prime}+\left(1+\delta_{\rm m}\right)\frac{\tilde{\theta}}{a} = &\; 0\;,\label{eqn:DMpt}\\
 \delta_{\rm de}^{\prime}+\frac{3}{a}\left(c_{\rm eff}^2-w_{\rm de}\right)\delta_{\rm de}+
 \left[1+w_{\rm de}+\left(1+c_{\rm eff}^2\right)\delta_{\rm de}\right]\frac{\tilde{\theta}}{a} = &\; 0 \;, 
 \label{eqn:DEpt}\\
 \tilde{\theta}^{\prime}+\left(\frac{2}{a}+\frac{H^{\prime}}{H}\right)\tilde{\theta}+\frac{\tilde{\theta}^2}{3a}+
 \frac{\tilde{\sigma}^2}{a}+
 \frac{3}{2a}\left[\Omega_{\rm m}\delta_{\rm m}+\Omega_{\rm de}\delta_{\rm de}\right] = &\; 0\;,\label{eqn:thetapt}
\end{align}

To determine the initial conditions necessary to solve the system of equations~\ref{eqn:DMpt} - \ref{eqn:thetapt} we 
refer to \cite{Pace2014b} and references therein for more details. Shortly, at early times the system will contain only 
linear terms in $\delta$ and $\theta$, therefore, also if we would have two different Euler equations, they will be 
identical in the linear regime. The initial overdensity $\delta_{\rm m,i}$ is chosen iteratively so that the 
overdensity at collapse time diverges. Once this quantity is known, and assuming a power-law behaviour at early times, 
$\delta_{\rm m}=Aa^n$, initial conditions for the dark energy component and the dimensionless peculiar velocity are
\begin{equation}
 \delta_{\rm de,i} = \frac{n}{\left[n+3\left(c_{\rm eff}^2-w_{\rm de}\right)\right]}(1+w_{\rm de})~\delta_{\rm m,i}\;,
 \quad \tilde{\theta}_{\rm i} = -n\delta_{\rm m,i}\;.
\end{equation}

The slope $n$ is of order unity for the EdS and the $\Lambda$CDM model and in general for all the models well 
approximated by an EdS model at early times. Deviations are appreciable but nevertheless still very small only for 
early dark energy models \citep{Ferreira1998,Batista2013}.

\section{Determination of the tidal shear}\label{sect:Tidal}
In this section we review the procedure to evaluate the tidal shear that will be included in the equations of the 
spherical collapse models (Eq.~\ref{eqn:thetaDMp}), as explained in detail in Sect.~\ref{sect:scm}.

To describe the effect of the surrounding shear on the halo we describe the motion of particles by the Zel'dovich 
approximation \citep{ZelDovich1970}:
\begin{equation}\label{eq:Zeld}
 x_i = q_i-D_+(t)\partial_i \psi \equiv q_i -D_+(t)\psi_{,i}\;.
\end{equation}
Here the displacement field $\psi$ is related to the density contrast $\delta$ via a Poisson relation, 
$\Delta \psi = \delta$. $D_+(t)$ is the linear growth factor. As Eq.~(\ref{eq:Zeld}) describes a potential flow, 
vorticity is naturally not generated and the only remaining object of the two invariants $\sigma^2$ and $\omega^2$ in 
Eq.~(\ref{eqn:thetaDMp}) is the traceless shear tensor which within this approximation reads
\begin{equation}\label{eq:shear}
 \sigma^2 \equiv \sigma_{ij}\sigma^{ij} =\dot{D}_+^2(t)\left(\psi_{,ij}\psi^{,ij}-\frac{1}{3}(\Delta\psi)^2\right)\;,
\end{equation}
with $\psi_{,ij}\equiv\partial_i\partial_j\psi$. With this Ansatz we describe external tidal fields acting on the 
collapsing halo in the linear theory only in a self-consistent way. This means that our treatment is restricted to 
fluctuations of the density field on large enough scales. Nonetheless this effect enters in the fully non-linear 
collapse equation.

Due to the Poisson equation we find in Fourier space
\begin{equation}
 \psi_{,ij} = \int\frac{\text{d}^3k}{(2\pi)^3}\frac{k_ik_j}{k^2}\delta(\boldsymbol k)\exp(\text{i}\boldsymbol{k}
              \boldsymbol{x})\;.
\end{equation}
In spherical coordinates this can be brought into the following form \citep[see][]{Regos1995,Heavens1999}:
\begin{equation}
 y^n_{lm} = \sqrt{4\pi}\frac{\text{i}^{l+2n}}{\sigma_{l+2n}}\int\frac{\text{d}^3k}{(2\pi)^3}k^{l+2n}
 \delta(\boldsymbol k) Y_{lm}(\hat{k})\exp(\text{i}\boldsymbol k\boldsymbol x)\;,
\end{equation}
with the direction vector $\hat{k}=\boldsymbol k/k$ and $\sigma_i$ being the spectral moments of the matter power 
spectrum
\begin{equation}
 \sigma_i^2 = \frac{1}{2\pi^2}\int\text{d}k\; k^{2i+2} P(k)\;,
\end{equation}
while $Y_{lm}$ are the spherical harmonics. The linear mapping \citep{Schaefer2012} between $y^n_{lm}$ 
and the tidal shear values $\psi_{,ij}$ is
\begin{equation}\label{eq:11}
 \begin{split}
  \sigma_0 y_{20}^{-1} = & \ -\sqrt{\frac{5}{4}}\left(\psi_{,xx}+\psi_{,yy}-2\psi_{,zz}\right)\;, \\
  \sigma_0 y^{-1}_{2\pm 1} = & \ -\sqrt{\frac{15}{2}}\left(\psi_{,xz}\pm\text{i}\psi_{,yz}\right)\;, \\
  \sigma_0 y^{-1}_{2\pm 2} = & \ \sqrt{\frac{15}{8}}\left(\psi_{,xx}-\psi_{,yy}\pm2\text{i}\psi_{,xy}\right)\;, \\
  \sigma_0 y^{0}_{00} = & \ \left(\psi_{,xx}+ \psi_{,yy}+\psi_{,zz}\right)\;,
 \end{split} 
\end{equation}
with the covariance matrix adopting the simple form:
\begin{equation}
 \left\langle y_{lm}^n(\boldsymbol x)y_{l'm'}^{n'}(\boldsymbol x)^*\right\rangle =(-1)^{n-n'}\frac{\sigma^2_{l+n+n'}}
 {\sigma_{l+2n}\sigma_{l+2n'}}\delta_{ll'}\delta_{mm'}\;.
\end{equation}
Thus, in the $y^n_{lm}$ basis the tidal shear values are uncorrelated Gaussian random variables with unit variance. We 
obtain the tidal shear values in physical coordinates by inverting the mapping.
The mass dependence of the tidal shear is represented by a low-pass filter acting on the power spectrum, thus cutting 
off high wavenumber modes:
\begin{equation}
 P(k) \to P(k)W^2_R(k), 
\end{equation}
with $W_R(k) = \exp(-k^2R^2/2)$. The mass scale is obtained via 
$M = \frac{4\pi}{3}\rho_\text{crit}\Omega_\text{m} R^3$, where $\rho_\text{crit}= 3H^2/(8\pi G)$ is the critical 
density. For more details we refer to \cite{Reischke2016a}.

\section{Evolution of \texorpdfstring{$\delta_{\rm c}$}{dc} in dark energy cosmologies}\label{sect:DEmodels}
One of the main quantities evaluated within the formalism of the spherical collapse model is the linear extrapolated 
density parameter $\delta_{\rm c}$, which represents the linear solution of the system of equations~\ref{eqn:DMpt} - 
\ref{eqn:thetapt}. This quantity is at the core of the theoretical framework used to evaluated the mass function, as we 
will see in Sect.~\ref{sect:mf}. Physically, this parameter represents the linear evolution of the primordial 
overdensity that gives rise to structures at a given epoch. In the standard approach, this quantity is a function of 
time only and it weakly depends on the background cosmological model. Differences might appear when also dark energy is 
clustering, but as shown for example in \cite{Batista2013}, $\delta_{\rm c}$ is closer to the $\Lambda$CDM model with 
respect to a smooth dark energy component. A higher (lower) value will result in a lower (higher) number of objects. 
In an extended approach, for example in the ellipsoidal collapse model \citep{Angrick2010} and in the extended 
spherical collapse model \citep{DelPopolo2013a,DelPopolo2013c,Pace2014b}, $\delta_{\rm c}$ depends also on mass.

In this section, extending the recent work by \cite{Reischke2016a}, we will consider the effects of tidal shear on 
clustering dark energy. An obvious comparison of our results will then be with \cite{Pace2014b}, but this can be done 
only at a qualitative level to discuss the general trend of $\delta_{\rm c}$ with mass and redshift: this is because in 
\cite{Pace2014b} both shear and rotation (with the latter dominating over the first) are included. Our main comparison 
therefore will be with our previous work \citep{Reischke2016a}. To do so, we will consider the following set of models, 
constituted by nine models, three with constant and six with a time-varying equation of state. Models with constant 
equation of state are the $\Lambda$CDM, a quintessence model with $w_{\rm de}=-0.9$ and a phantom model with 
$w_{\rm de}=-1.1$. Note that the last two models are excluded by recent CMB observations \citep{Planck2016_XIII}, but 
here we will consider them merely for the sake of comparison over a variety of models. Note also that if we want 
$w_{\rm de}\neq -1$, but still within the observational limits, perturbations in dark energy will be very small, being 
$\delta_{\rm de}\propto (1+w_{\rm de})\delta_{\rm m}$. Dynamical models are:
\begin{itemize}
 \item the 2EXP model \citep{Barreiro2000},
 \item the CNR and the SUGRA model \citep{Copeland2000},
 \item the CPL model \citep{Chevallier2001,Linder2003},
 \item the INV1 and INV2 models \citep{Corasaniti2003,Corasaniti2004,Sanchez2009}.
\end{itemize}
The $\Lambda$CDM model, albeit not being affected by perturbations in the dark energy sector, will be our reference 
model.

We refer to \cite{Szydlowski2006} and \cite{Jennings2010} for a detailed discussion of their properties and time 
evolution. Here we limit ourselves to present their functional form and parameters which we will describe in the 
following. 
The CPL model is described by the following linear evolution with respect to the scale factor:
\begin{equation}
 w_{\rm de}(a)=w_0+w_\text{a}(1-a)\;,
\end{equation}
and we used $w_0=-1$ and $w_{\rm a}=0.15$. Note that the slope $w_{\rm a}$ is much less gentle of what allowed by 
\cite{Planck2016_XIII,Planck2016_XIV} \citep[but see also][for a recent use in the reconstruction of the potential of 
scalar field models]{Battye2016}.

The other models are described by the following functional form
\begin{equation}
 w_{\rm de}=w_0+(w_{\rm m}-w_0)\frac{1+e^{\frac{a_{\rm m}}{\Delta_{\rm m}}}}
 {1+e^{-\frac{a-a_{\rm m}}{\Delta_{\rm m}}}}
 \frac{1-e^{-\frac{a-1}{\Delta_{\rm m}}}}{1-e^{\frac{1}{\Delta_{\rm m}}}}\;,
\end{equation}
and their parameters are presented in \autoref{tab:params}.

\begin{table}
 \caption{Parameter values for the dark energy models with dynamical equation-of-state parameter.}
 \begin{center}
  \begin{tabular}{c|c|c|c|c|c|}
   \hline
   \hline
   Model & $w_0$ & $w_{\mathrm{m}}$ & $a_{\mathrm{m}}$ & $\Delta_{\mathrm{m}}$ \\
   \hline
   2EXP  & -0.99 & 0.01 & 0.19 & 0.043 \\
   INV1  & -0.99 & -0.27 & 0.18 & 0.5 \\
   INV2  & -0.99 & -0.67 & 0.29 & 0.4 \\
   CNR   & -1.0 & 0.1 & 0.15 & 0.016 \\
   SUGRA & -0.99 & -0.18 & 0.1 & 0.7 \\
   \hline
  \end{tabular}
 \end{center}
 \label{tab:params}
\end{table}

We will also assume the following cosmological parameters:
$\Omega_{\rm m}=0.32$, $\Omega_{\rm de}=0.68$, $h=0.67$ and $n_{\rm s}=0.966$.

Once a model is selected, it is possible to evaluate $\delta_{\rm c}$. To do so we proceed as follows:
\begin{itemize}
 \item Fix the collapse redshift
 \item Fix the mass of the overdensity and select the corresponding value of the shear from a pre-computed table
 \item Find the initial conditions as explained in Sect.~\ref{sect:scm} for the particular choice of mass and redshift
 \item Solve the linearised version of equations \ref{eqn:DMpt} - \ref{eqn:thetapt} with the initial conditions 
       determined above
\end{itemize}

\begin{figure*}
 \centering
 \includegraphics[width=0.3\textwidth,angle=-90]{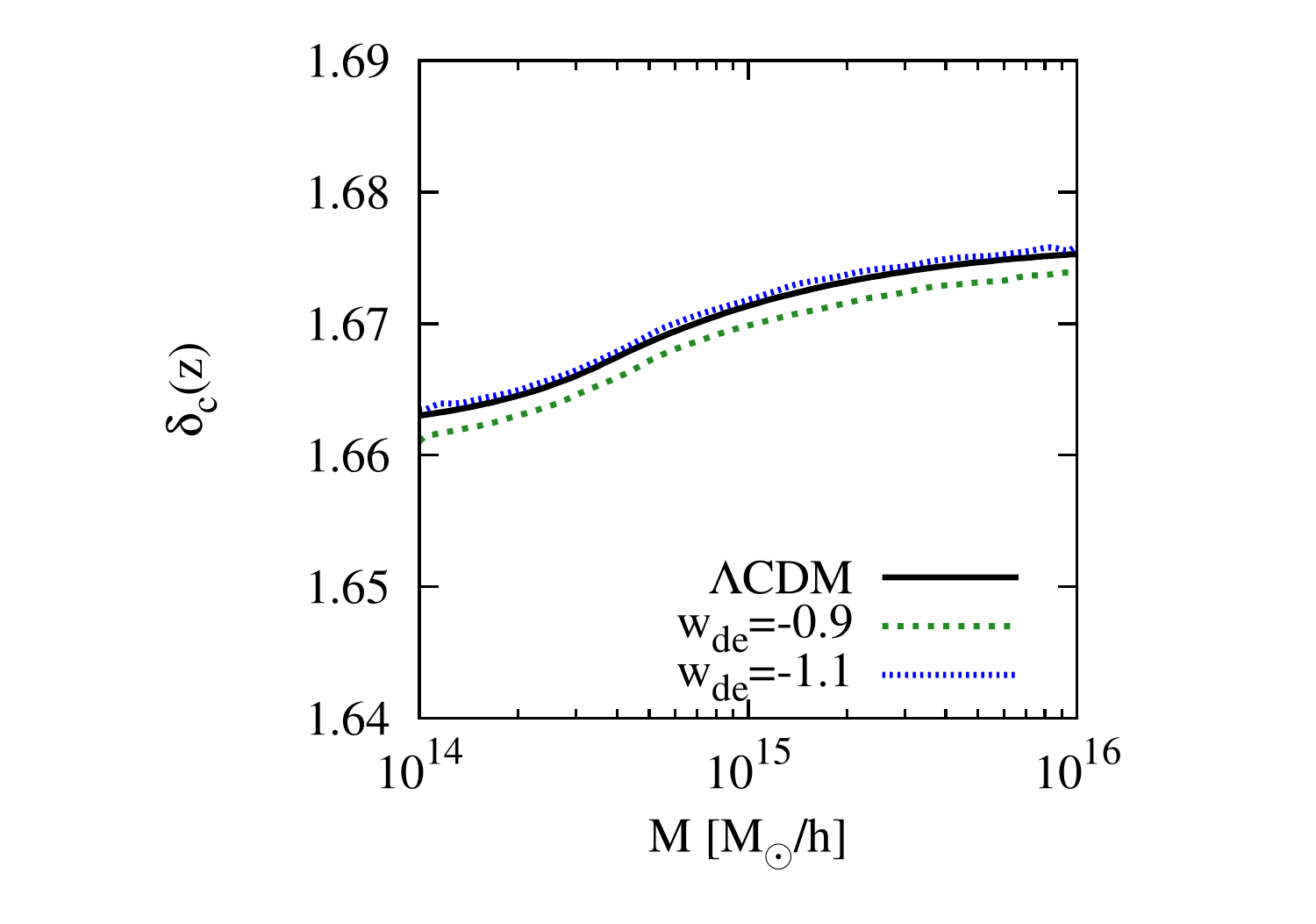}
 \includegraphics[width=0.3\textwidth,angle=-90]{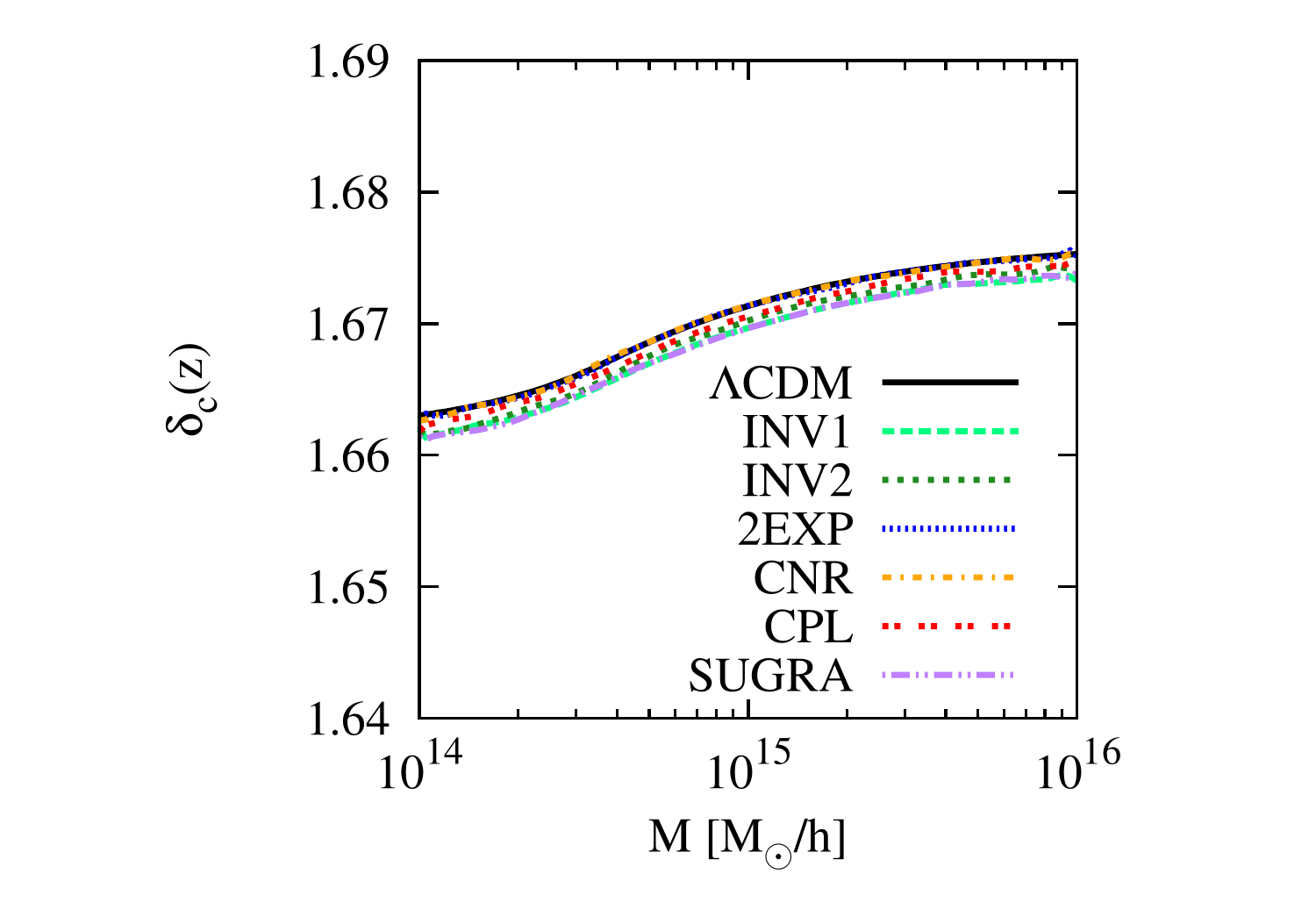}\\
 \includegraphics[width=0.3\textwidth,angle=-90]{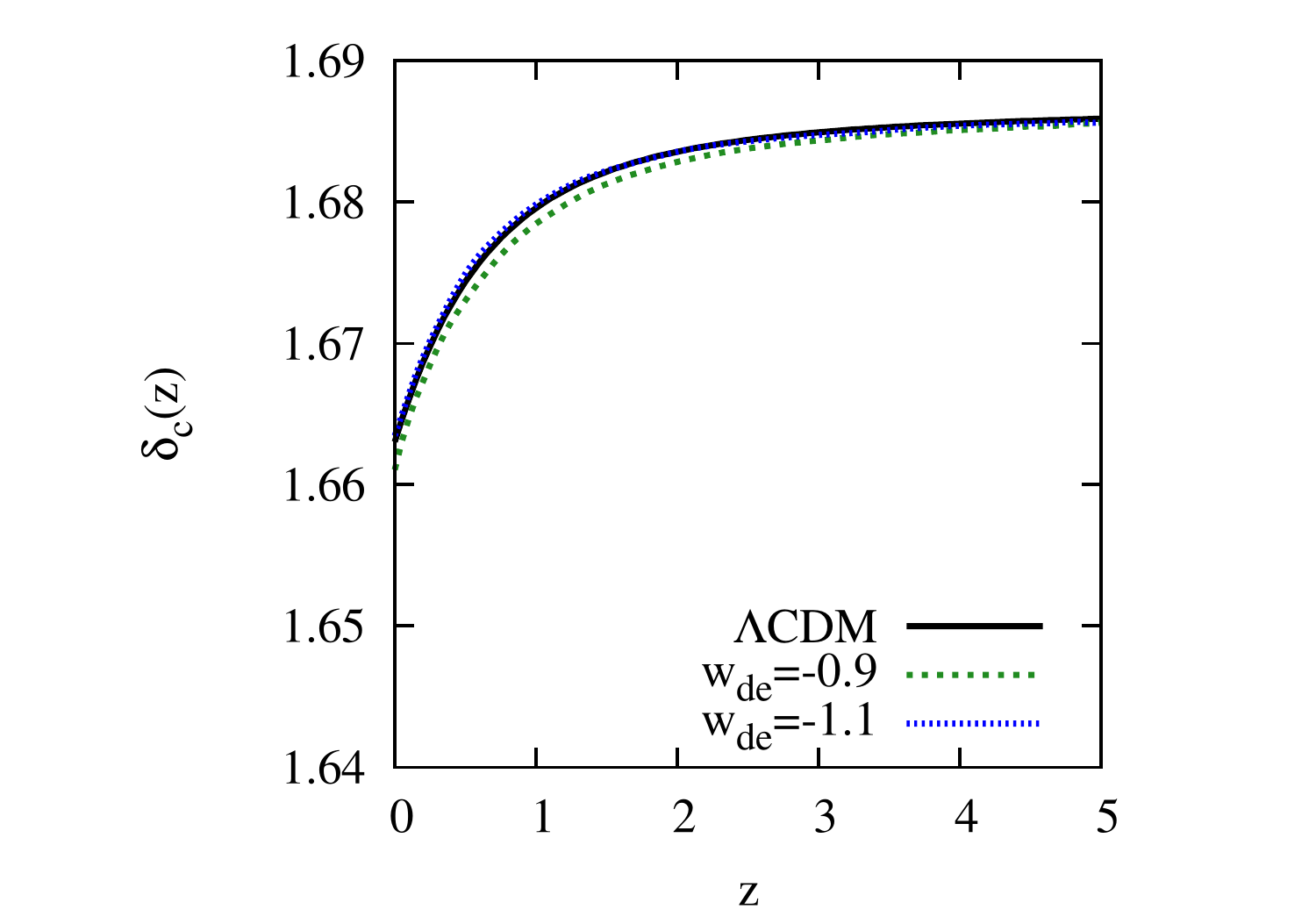}
 \includegraphics[width=0.3\textwidth,angle=-90]{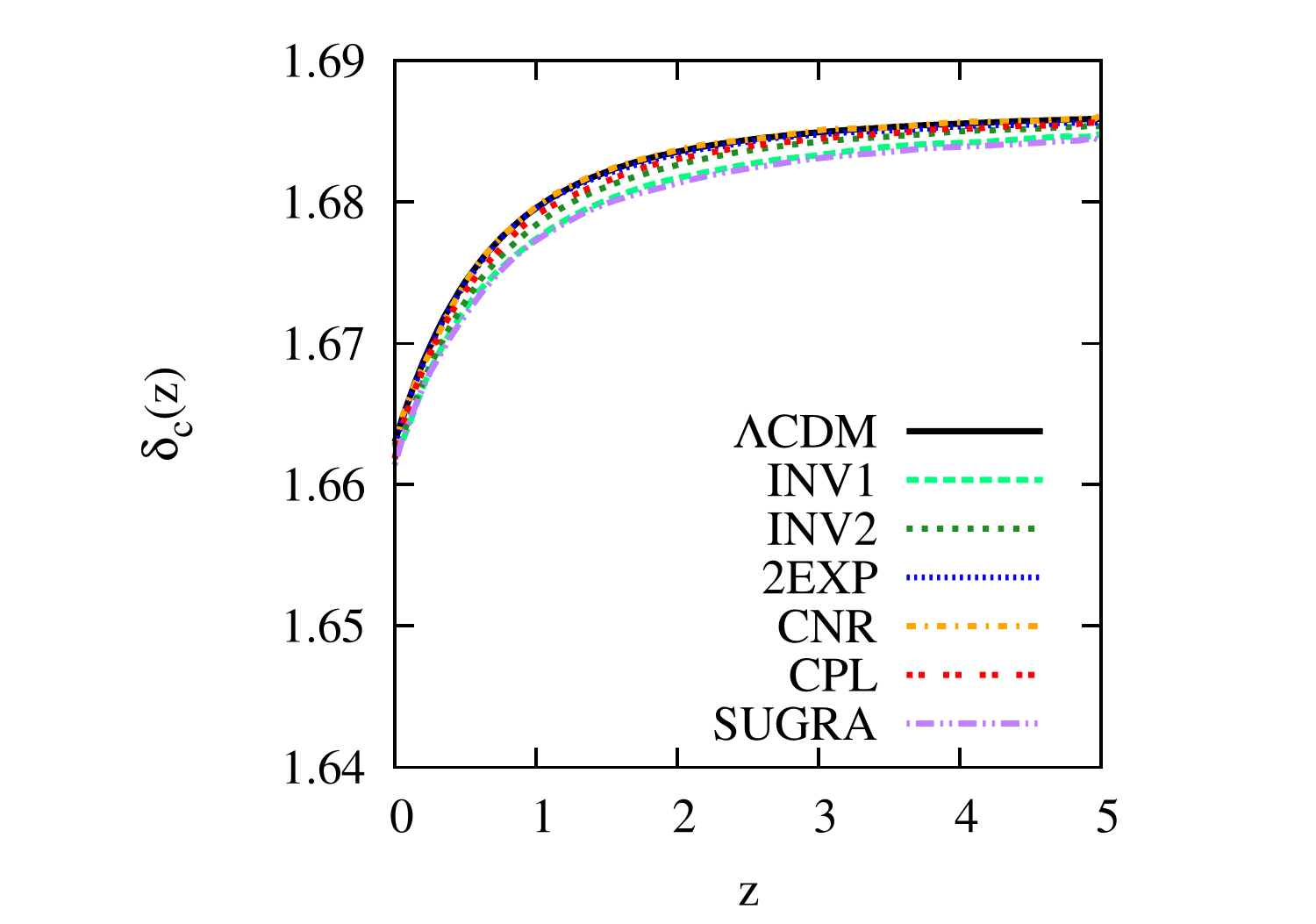}
 \includegraphics[width=0.3\textwidth,angle=-90]{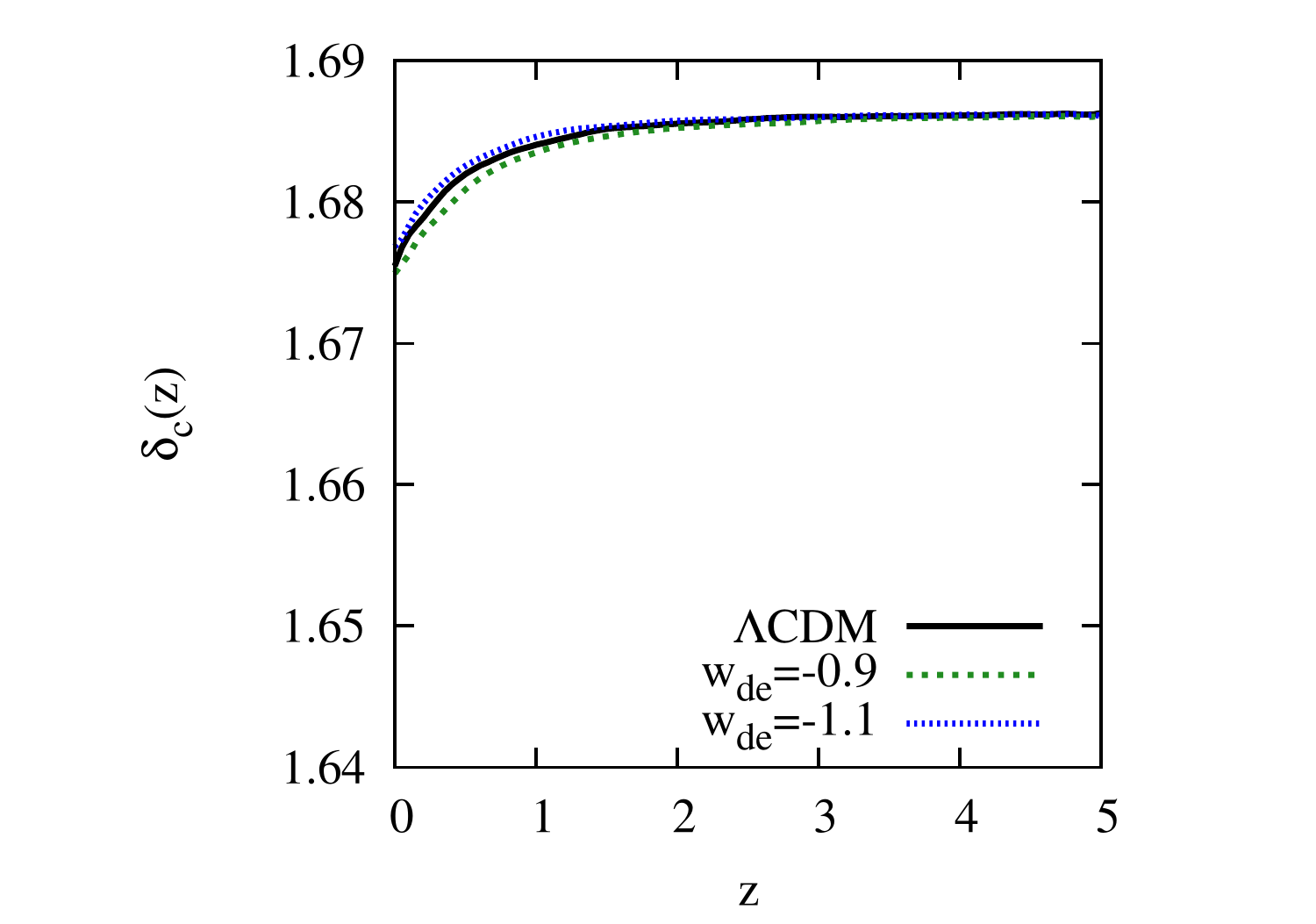}
 \includegraphics[width=0.3\textwidth,angle=-90]{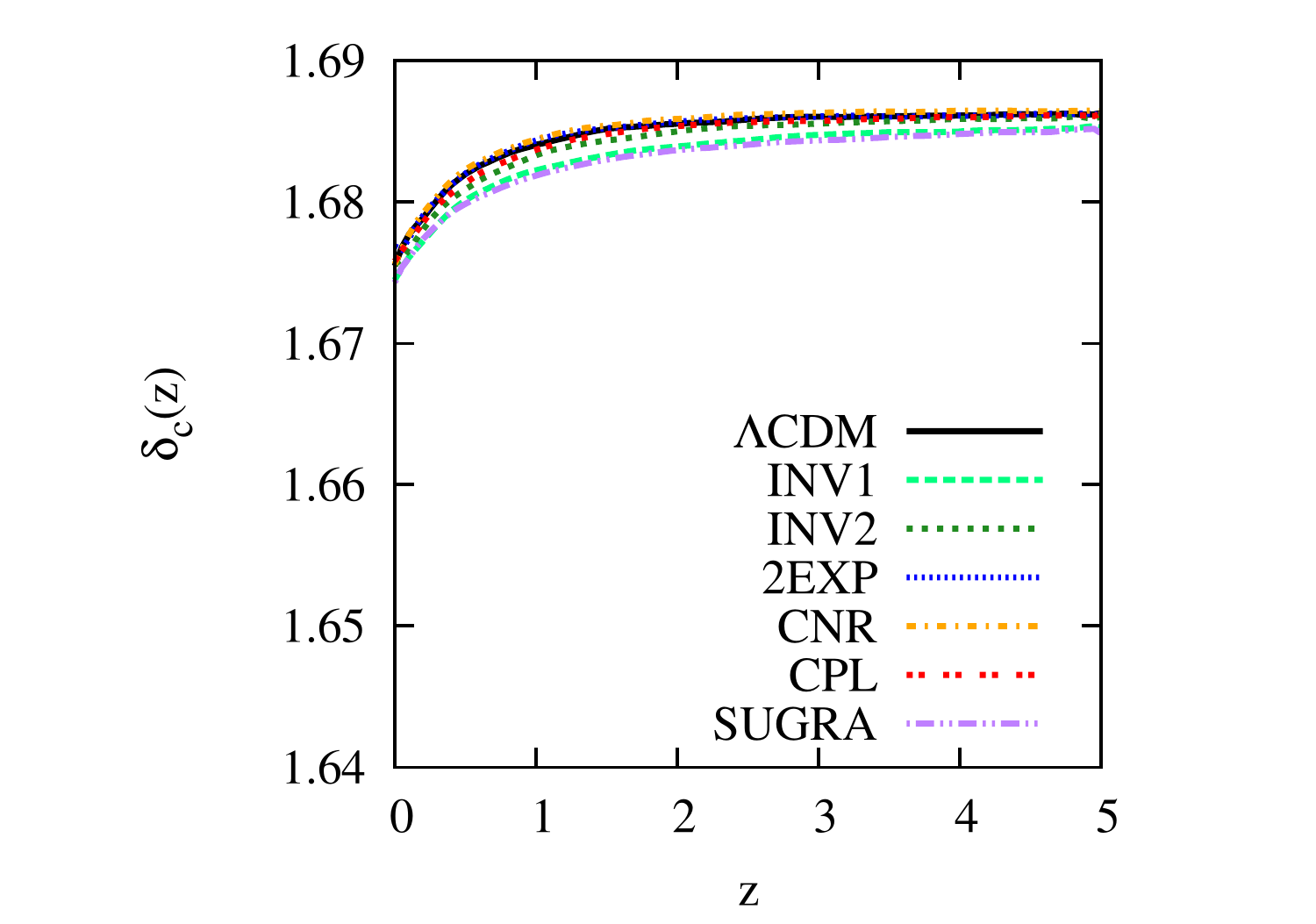}
 \caption{\textit{Upper panels}: effects of the tidal shear on $\delta_{\rm c}$ at $z=0$ for different values of the 
 mass of the collapsing sphere. \textit{Middle panels}: time evolution of $\delta_{\rm c}$ for the minimum mass 
 sampled ($M=10^{14}~M_{\odot}/h$). \textit{Bottom panels}: time evolution of $\delta_{\rm c}$ without the inclusion of 
 the tidal shear. Left (right) panels refer to constant (dynamical) equations of state. The black solid line refers to 
 the reference $\Lambda$CDM model. For models with constant equation of state, the green short-dashed (blue dotted) 
 curve shows a quintessence (phantom) model with $w_{\rm de}=-0.9$ ($w_{\rm de}=-1.1$). 
 For dynamical dark energy models, the light green dashed (dark green short-dashed) curve represents the INV1 (INV2) 
 model; the blue dotted curve the 2EXP model; the CPL (CNR) model with the red dashed-dot (orange dot-dashed) curve 
 and finally the SUGRA model with the violet dashed-dot-dotted curve.}
 \label{fig:deltac}
\end{figure*}

In \autoref{fig:deltac} we present the evolution of the linear extrapolated density parameter $\delta_{\rm c}$ with 
(top panels) and without (bottom panels) the inclusion of the tidal shear, taking as reference model the $\Lambda$CDM 
model (black solid line). Left (right) panels show models with a constant (dynamical) equation of state. 
In the caption we describe the different colours and line-styles of the models analysed in this work.

As already described in \cite{Reischke2016a} we restrict our treatment to masses above $10^{14}\; M_\odot$ as smaller 
masses would lead to high values of $\delta = \psi_{,ii}$ as the cut-off of the power spectrum would be at higher $k$. 
In this regime the treatment of the Zel'dovich approximation used to calculate the shear would break down and higher 
order effects which cannot be related to properties of the statistics of the underlying density field in a clear way 
need to be taken into account. Nonetheless the expected trend for smaller masses is an increase of the importance of 
tidal fields as the local fluctuations of the smoothed density field become larger.

As it appears clear from the top and middle rows of \autoref{fig:deltac} after a direct comparison with the 
corresponding panels in Fig.~7 of \cite{Reischke2016a}, clustering dark energy models with the inclusion of the tidal 
shear $\sigma$ resemble more closely the $\Lambda$CDM model at all redshifts and at all mass scales with respect to the 
corresponding smooth models. This feature has been already observed in \cite{Batista2013} for early dark energy models 
and we refer the reader to this work for a detailed explanation.

As expected, by construction, tidal shear has a bigger influence on small mass scales ($M\approx 10^{14}~M_{\odot}/h$). 
The linear extrapolated linear overdensity $\delta_{\rm c}$ increases from 
$\approx 1.665$ ($M\approx 10^{14}~M_{\odot}/h$) to $\approx 1.675$ ($M\approx 10^{16}~M_{\odot}/h$). In the high mass 
regime we notice a flattening of the curve, as in \cite{Reischke2016a}, due to the fact that for these masses the 
effect of the tidal shear is completely negligible.

Note once again how differences between the $\Lambda$CDM model and the dark energy models are minimal. By fixing the 
mass at $M=10^{14}~M_{\odot}/h$, where we expect the biggest effect, we can also study the time evolution of 
$\delta_{\rm c}$ with respect to the redshift $z$. 
As explained in \cite{Reischke2016a} and in Sect.~\ref{sect:Tidal}, the tidal shear is a function of both mass and 
time, with bigger effects at small redshifts, as it also happens for the $\alpha$ parameter (defined via 
$-\frac{3}{2}\alpha(\Omega_{\rm m}\delta_{\rm m}+\Omega_{\rm de}\delta_{\rm de})=(\sigma^2-\omega^2)/H^2$) in 
\cite{Pace2014b}. At high redshifts all the models recover very well an EdS universe since the effect of the tidal 
shear is negligible. At small redshifts all the models present slightly smaller values with respect to the case when 
the tidal shear is not included.

These results have the same qualitative behaviour of \cite{Pace2014b}: bigger effect at small masses and redshifts and 
negligible effect at high masses and redshifts. Here nevertheless the effect is much smaller and $\delta_{\rm c}$ is 
lower than for the standard spherical collapse model while the opposite takes place when the contribution of the 
angular momentum is included.

In general, we notice that the clustering of dark energy has negligible effects on the tidal shear as the equation of 
state had for the smooth case. This is due to the fact that tidal shear effects are important only at late times 
but theoretically tidal shear is evaluated at early times and as shown in \cite{Reischke2016a} this gives a per mill 
deviation from the standard spherical collapse model. In addition, at early times, all the models are approximately the 
same (dark energy perturbations and energy density are completely negligible with respect to the dark matter component) 
therefore the shear term is identical for all the models, as we verified numerically. This implies that the non-linear 
term due to shear is only a very small correction to the dynamical equations describing the spherical evolution.

The case presented and discussed here assumes that both dark matter and dark energy are affected in the same way by the 
tidal shear. We also studied a more general case where the tidal shear only affects the equations of motion of the dark 
matter component but not those of dark energy. Due to the above considerations, we could expect analogous results to 
before, as a direct inspection confirmed. Results are not only in qualitative, but also in quantitative agreement to 
what already discussed. The reason is due to that fact that linearised equations at early times are identical in both 
cases (by definition) and that the tidal shear is a very small contribution at such high redshifts. 
We will therefore not distinguish the two cases any more in the following. 
We also remark that the lack of differences between the two hypotheses was also found in \cite{Pace2014b}.

\section{Mass function and number counts}\label{sect:mf}
Within the framework of the Press \& Schechter \citep{Press1974} and of the Sheth \& Tormen \citep{Sheth2001,Sheth2002} 
formalism, knowing the time evolution of the growth factor (which describes the linear evolution of structures) and of 
the linear overdensity parameter $\delta_{\rm c}$, it is possible to evaluate the abundance of halos. A further 
parameter that needs to be specified is the normalization of the linear matter power spectrum $\sigma_8$. A theoretical 
derivation for the expression of the mass function within the theory of Gaussian random fields can be found in 
\cite{Press1974} and subsequently within the formalism of the excursion set \citep{Bond1991}. Due to its problems at 
both the low and high mass tail of the mass function, we decided to use the functional form provided in 
\cite{Sheth1999,Sheth2001,Sheth2002}. Despite the fact that this improved formulation based on the ellipsoidal collapse 
model fits better the simulations at $z=0$, it still has problems at high redshifts, where it significantly 
overestimates the number density of halos \citep{Klypin2011}. Limiting ourselves to low redshifts, we will use the 
Sheth \& Tormen mass function formulation. 
We will also assume that this formulation can be safely extended to clustering dark energy models 
\citep[but see also][]{DelPopolo1999,DelPopolo2006}.

Before discussing in detail our finding, there is a further point we would like to clarify. The mass function and 
quantities derived, such has the number of halos above a given mass or in a range of redshifts, depend on the mass, but 
for clustering dark energy models there is an ambiguity in the definition of the mass to be used. When dark energy is 
smooth, the halo mass $M_{\rm h}$, assuming spherically symmetry and a top-hat profile as usually done within the 
formalism of the spherical collapse model, is $M_{\rm h}=\frac{4}{3}\pi\rho_{\rm h}R_{\rm h}^3$, where $R_{\rm h}$ is 
the halo virial radius and $\rho_{\rm h}=\bar{\rho}(1+\delta)$, where $\bar{\rho}$ is the background matter density and 
$\delta$ the overdensity. 
When dark energy can cluster instead, in principle the mass of the halo should be influenced by the dark energy 
perturbations and these should be taken into account. Since different authors define this correction in different ways 
\citep{Creminelli2010,Batista2013} and the contribution is nevertheless small \citep{Basse2012} and here we are 
interested to the interplay of the tidal shear and the effective sound speed $c_{\rm eff}^2$, we will not discuss this 
point further in our analysis.

Finally, a careful reader may wonder whether it is acceptable to use the Sheth \& Tormen mass function also for 
clustering dark energy models. To our best knowledge, we think this is a good assumption for several reasons: to-date, 
only \cite{Press1974}, \cite{Sheth2001} and \cite{DelPopolo2006} proposed theoretically motivated formulations for the 
mass function, while all other analytic expressions \citep[see][for a recent review]{Murray2013a,Murray2013b} are 
fitting formulae to N-body simulations. Moreover, with the exception of \cite{Bhattacharya2011}, all the others mass 
functions described in literature are tested again $\Lambda$CDM cosmologies only. It would therefore be hard to know 
whether the fitted parameters would be changed in dark energy or clustering dark energy models. In addition, so far, 
there are no simulations allowing to study clustering dark energy in the fluid formalism and since all of the proposed 
mass functions agree reasonably well at $z=0$ and differ, due to a low number of objects, only at high mass, we 
conclude that the use of the mass function proposed in \cite{Sheth2001} is legit.

In the mass function formalism here adopted, the main theoretical quantity entering into the expression for the mass 
function is $\nu=\delta_{\rm c}/\sigma$, where $\delta_{\rm c}$ was discussed in the previous section and $\sigma$ 
represents the evolution in time of the mass variance of the linear matter power spectrum. This quantity evolves 
proportionally to the linear growth factor, which is affected by cosmology only but not by the tidal shear. 
Differently from the standard spherical collapse model, now $\delta_{\rm c}$ is also a function of mass and therefore 
we can expect a different shape of the mass function. This was also noticed and discussed in detail by 
\cite{Reischke2016a}. In addition we recall that the tidal shear evolves differently from the linear growth factor, 
therefore the time evolution of $\nu$ will be affected as well and also this will have consequences on the overall mass 
function.

In figure~\ref{fig:mf} we show the cumulative comoving number density of objects above a given mass at $z=0$; in this 
way our analysis will not be affected by volume effects. In addition, we will also assume that all models have the same 
normalization of the matter power spectrum. Since from the above analysis it is clear that differences will be small, 
we prefer not to hide them behind normalization effects. For a better understanding of our findings we show the ratio 
between the dark energy and the $\Lambda$CDM model. Left panels show the ratio between the models with and without 
tidal shear contribution while right panels show the ratio between the dark energy and the $\Lambda$CDM model with 
tidal shear field.

The cumulative number density of objects above a given mass is defined as
\begin{equation}
 n(>M) = \int_{M}^{\infty}dM^{\prime}\frac{dn}{dM^{\prime}}(M^{\prime},z)\;,
\end{equation}
where $\frac{dn}{dM}(z)$ is the mass function evaluated at redshift $z$. 
The minimum mass used is $10^{14}~h^{-1}M_{\odot}$.

As a direct consequence of the analysis on the spherical collapse model, differences on the comoving number density are 
small. Smallest effects take place at low masses ($M\approx 10^{14}~M_{\odot}/h$) and {differences increase towards 
higher masses. At the highest mass probed here, differences are of the order of 10\%, in perfect analogy with 
the case of smooth dark energy. Differences between the dark energy models and the reference $\Lambda$CDM model are 
very small and noticeable mainly at high masses. This is in agreement with the fact that clustering dark energy models 
are more similar to the $\Lambda$CDM model with respect to a smooth model. Differences are at most 3\% for the 2EXP 
model. 
The top right panel shows models with a constant equation of state and we notice that both the quintessence and phantom 
models have a very similar behaviour, with the latter in particular very close to the $\Lambda$CDM model. 
In the bottom right panel we show the same quantity, but for dynamical dark energy models. Once again the models behave 
very similarly to each other. Major effects take place for the 2EXP model while other models, such as the CPL and the 
INV1 and INV2 are practically indistinguishable from the $\Lambda$CDM model.

Having established the relative effect of the tidal shear for each model, we now want to study the ratio between the 
dark energy models and the $\Lambda$CDM one, when the tidal shear field is included for both. Results are in perfect 
qualitative agreement with what found in \cite{Reischke2016a} for smooth dark energy models. Starting with models with 
constant equation of state, quintessence (phantom) models predict more (less) objects with respect to the $\Lambda$CDM 
model. Differences take place only at high masses and they are of the order of 10\% (5\%) for quintessence (phantom) 
models. These differences are very similar to the equivalent smooth models, differing by only 2-3\%.

Qualitatively similar to the smooth scenario are also the dynamical dark energy models. Once again, these models, being 
in the quintessence regime, predict more objects than the $\Lambda$CDM model and the relative number of objects grow 
with mass. Also for this set of models results are in qualitative agreement with the smooth scenario, albeit 
quantitative differences with respect to the $\Lambda$CDM model are more limited. This is, once again, due to the fact 
that differences coming from $\delta_{\rm c}$ are more limited. Models with higher differences are the INV1 and SUGRA, 
while the CNR and the 2EXP basically predict the same number of objects of the $\Lambda$CDM model. Other models (INV2 
and CPL) show intermediate values.

In \cite{Reischke2016a}, the authors discuss the effect of tidal shear on the differential mass function and showed 
that its contribution is not enough to make the Press \& Schechter mass function closer to the Sheth \& Tormen one. 
Since also for clustering dark energy effects due to the inclusion of the tidal shear are small, we verified that a 
similar behaviour holds.

\begin{figure*}
 \centering
 \includegraphics[width=0.33\textwidth,angle=-90]{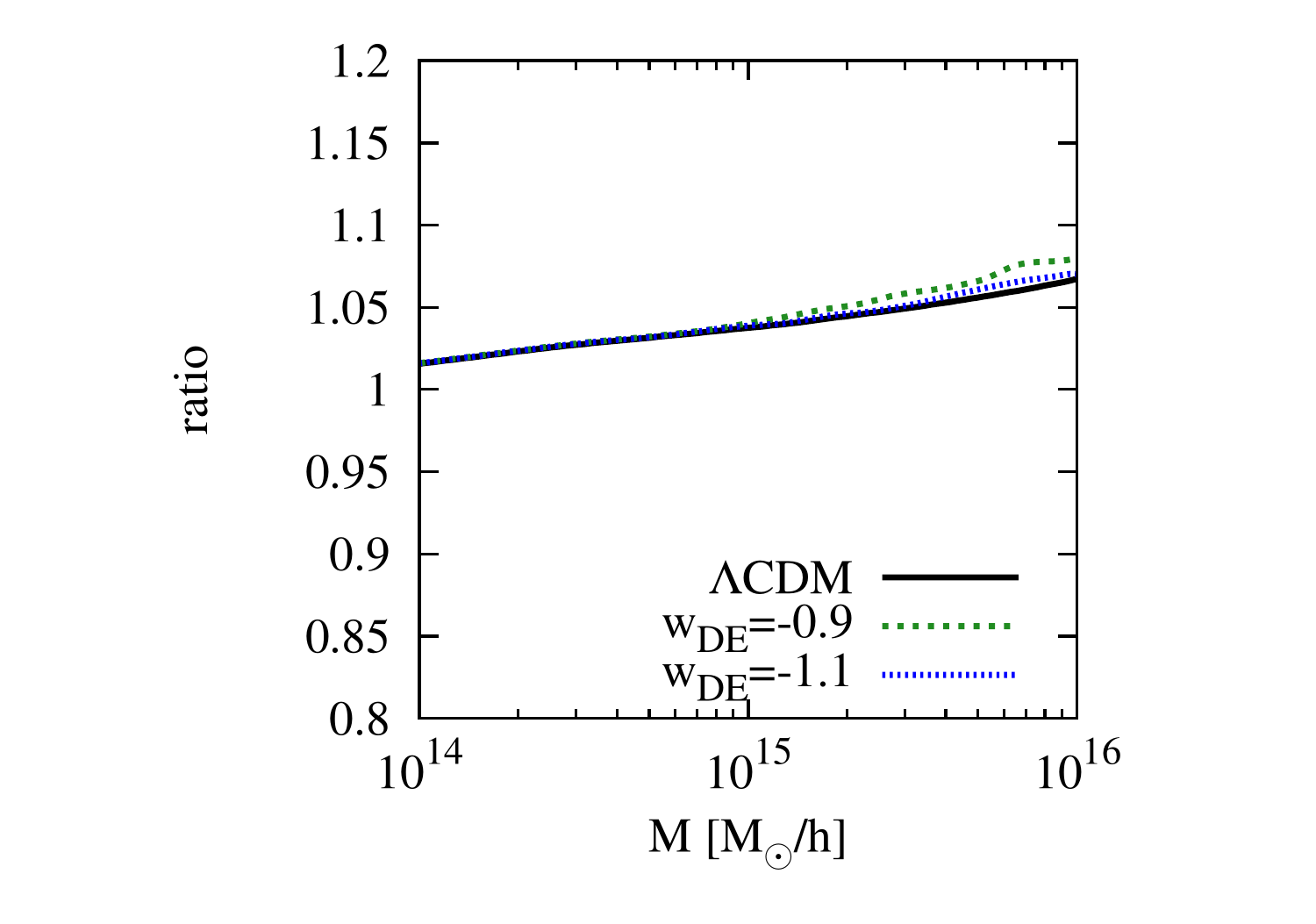}
 \includegraphics[width=0.33\textwidth,angle=-90]{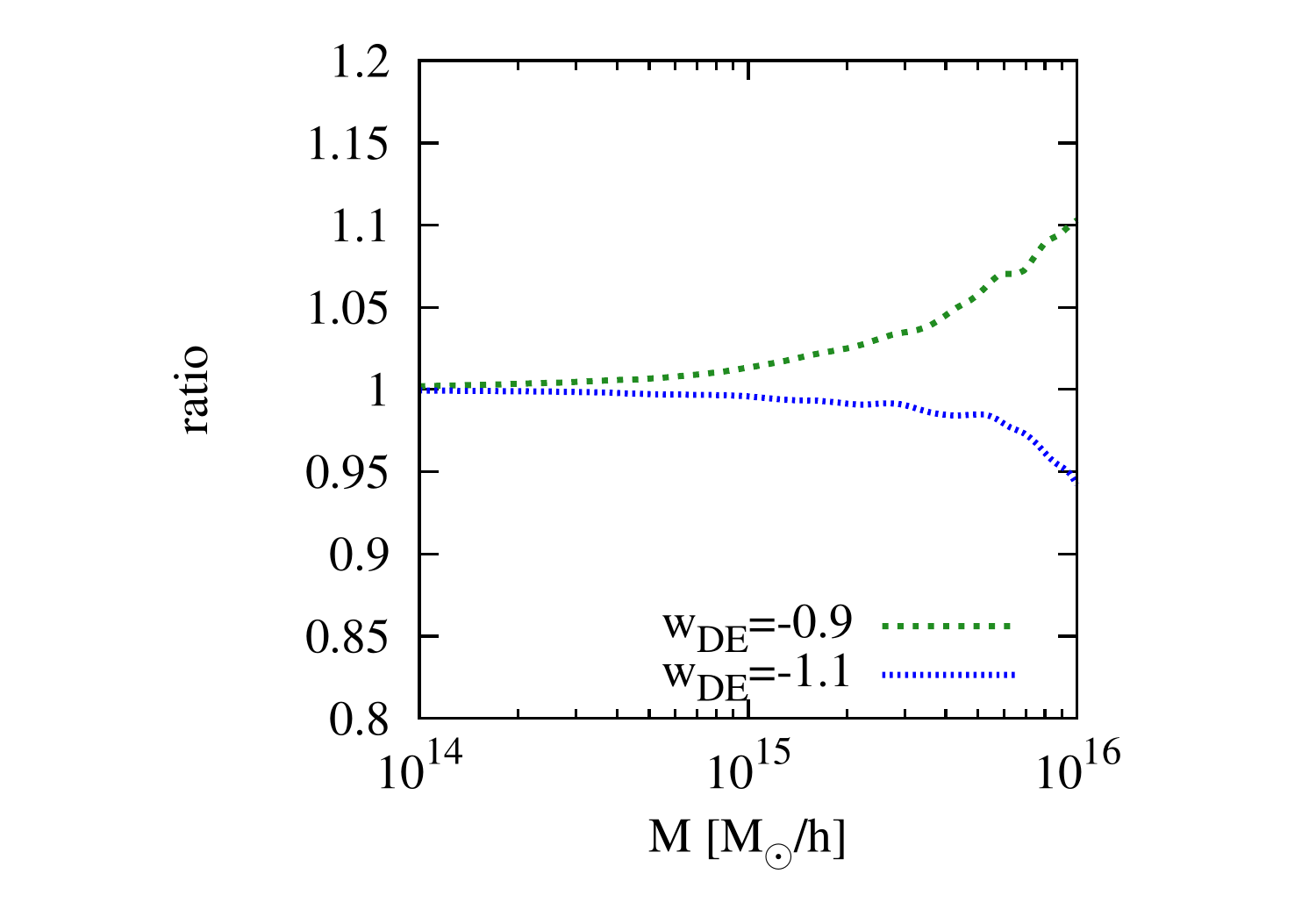}\\
 \includegraphics[width=0.33\textwidth,angle=-90]{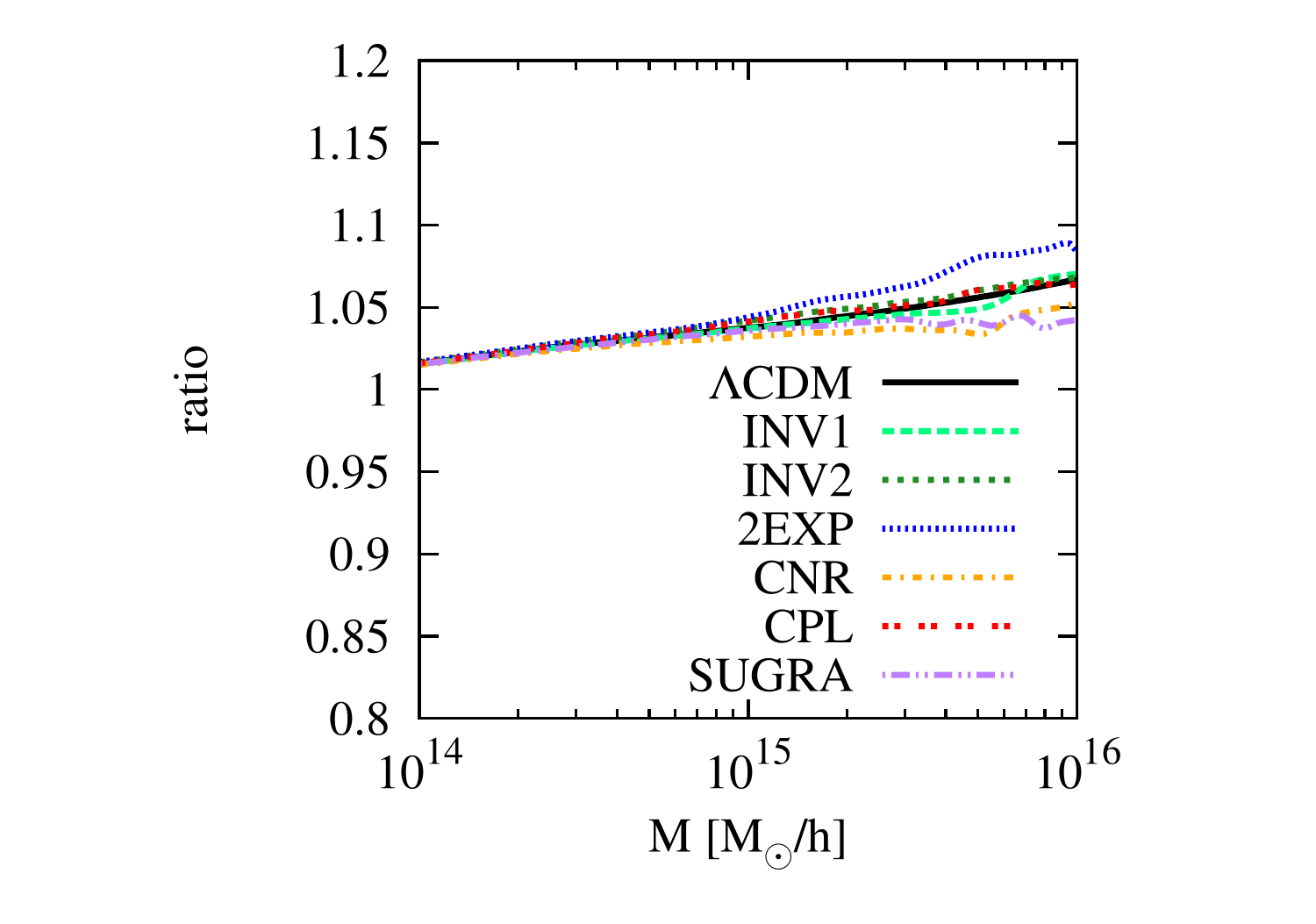}
 \includegraphics[width=0.33\textwidth,angle=-90]{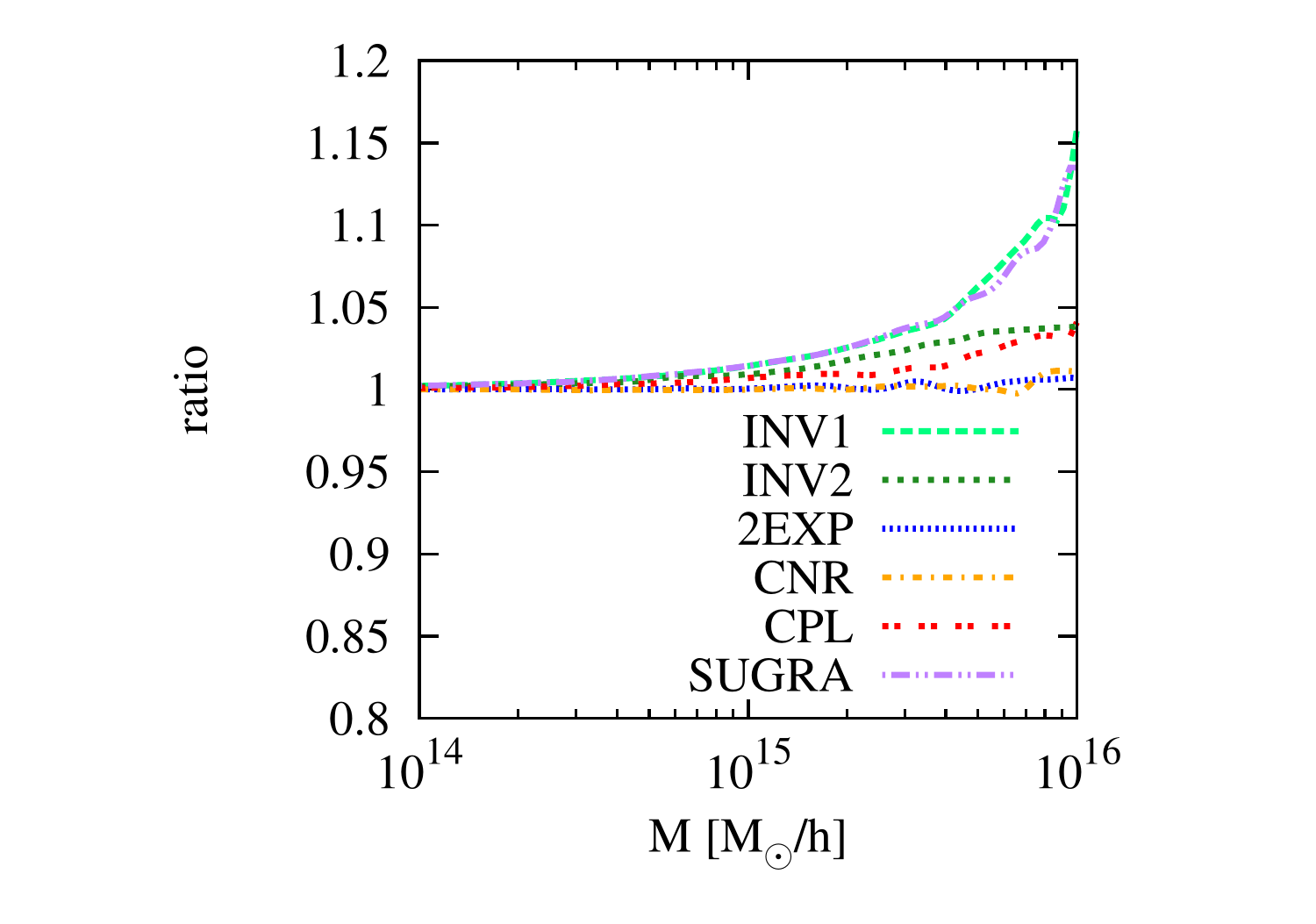}
 \caption{Ratio of the cumulative comoving number density of objects above mass $M$ evaluated at $z=0$. 
 \textit{Left column}: ratio between the number counts of the models with and without tidal shear. 
 \textit{Right column}: ratio between the dark energy and the $\Lambda$CDM model taking into account the tidal shear. 
 Upper panels show results for models with constant equation of state ($w_{\rm DE}=-0.9$ and $w_{\rm DE}=-1.1$). Lower 
 panels instead show results for dynamical dark energy models. Line-styles and colours are as in \autoref{fig:deltac}.}
 \label{fig:mf}
\end{figure*}

The effective sound speed of dark energy perturbations can affect several observational probes, such as the growth of 
matter perturbations, redshift space distortions and the shape of the dark matter power spectrum. These quantities will 
affect the weak lensing and the galaxy power spectrum which will be measured to a high level of accuracy by future 
surveys as Euclid \citep{Amendola2016}. 
The observational probes mentioned will depend on the quantity $Q$ which describes deviations from the standard Poisson 
equation: when dark energy is smooth (clustering), $Q=1$ ($Q\neq 1$). Assuming a null anisotropic stress, perturbations 
are fully described by $Q$ which can be parametrised as \citep[assuming a constant equation of state,][]{Sapone2009}
\begin{equation}
 Q(k,a) = 1+\frac{1-\Omega_{\rm m,0}}{\Omega_{\rm m,0}}\frac{(1+w)a^{-3w}}{1-3w+\frac{2}{3}\nu(a)^2}\;,
\end{equation}
where $\Omega_{\rm m,0}$ is the matter density parameter today and $\nu(a)\equiv c_{\rm eff}k/(aH)$. This expression is 
valid for both super- and sub-horizon scales. On scales inside the sound horizon, $Q$ is sensitive to $c_{\rm eff}$ 
when dark energy is far from being smooth, i.e. $c_{\rm eff}^2\lesssim 10^{-2}$. In other words, if dark energy is 
given by a minimally coupled scalar field, then its perturbations will be so small that we can neglect them.

Thanks to exquisite measurements in the weak and strong lensing regime, Euclid will be able to measure the halo mass 
function down to $10^{11}~h^{-1}M_{\odot}$ and therefore cover the mass regime studied in this work. 
Unfortunately, at these small scales it is necessary also to take into account effects from baryon physics and to 
understand the bias between the luminous and the dark matter component. Since these uncertainties can easily overcome 
the effect of tidal shear on the mass function, it might difficult to probe its effect. 
Also take into account that as shown in \cite{Reischke2016a}, the error on cosmological parameters when not taking 
into account the tidal shear can be up to 1\%. Since bigger effects will take place if $c_{\rm eff}^2\ll 1$, but at 
the same time the models will be closer to a $\Lambda$CDM model than the corresponding smooth counterpart, it is 
necessary a combination of probes to disentangle the opposing physical effects.

\section{Conclusions}\label{sect:discussion}
By using the Zel'dovich approximation to sample the external tidal shear, we extend the work by \cite{Reischke2016a} to 
clustering dark energy models. In our formalism it is not possible to include the effects of rotation and therefore a 
direct comparison with \cite{Pace2014b} is not possible, since there the main contribution is given by the rotational 
term. Here we limit ourselves to a qualitative comparison.

As for the case of smooth dark energy, tidal shear effects are more relevant for small masses and for lower redshifts 
since with the growth of structures the curvature of the potential increases.

With respect to the standard spherical case and in analogy to the smooth dark energy model, tidal shear supports the 
collapse. Its effect on the spherical collapse parameter $\delta_{\rm c}$ is of the order of few percent for the 
$\Lambda$CDM model and for the clustering dark energy models analysed in this work. We notice that also when the tidal 
shear is included, clustering dark energy models resemble the $\Lambda$CDM model more closely than a smooth dark energy 
model, as already explained for the standard spherical collapse model by \cite{Batista2013}.

Tidal shear affects the cumulative comoving number density of halos by few percent and major differences (of the order 
of 10\%) arise at high masses. Similar effects, both at a qualitative and at a quantitative level, appear when 
comparing dark energy models with the $\Lambda$CDM model. 
Comparing our results with \cite{Reischke2016a}, we show that the clustering of dark energy is largely unaffected 
by tidal shear and results agree quantitatively with the smooth case. 
In \cite{Pace2014b}, extending the work of \cite{DelPopolo2013a}, the effect of shear and rotation was studied in 
clustering dark energy models. A direct comparison can not be performed since in these previous works the dominant term 
is the rotation one, therefore we can only discuss analogies and differences at a qualitative level. Both approaches 
lead to a mass-dependent term, the collapse is favoured and effects are at the percent level, differently from the 
aforementioned works. The effects of the additional rotation term will be included in a following work.

Several probes can be used to measure the effective sound speed of dark energy perturbations $c_{\rm eff}^2$, but 
these are sensitive to it only if $c_{\rm eff}^2\ll 1$. Weak and strong lensing will be used to constrain the halo mass 
function and, as shown, effects will be appreciable in the case of fully clustering dark energy. On the other side, the 
linear extrapolated overdensity $\delta_{\rm c}$ will be closer to the $\Lambda$CDM model with respect to the smooth 
counterpart model, making therefore more difficult to measure the combined effects.

\section*{Acknowledgements}
The authors thank an anonymous referee whose comments helped to improved the scientific content of this work. 
FP acknowledges support from a post-doctoral STFC fellowship and thanks Inga Cebotaru for carefully reading the 
manuscript and improve the style. RR acknowledges funding by the graduate college Astrophysics of cosmological probes 
of gravity by Landesgraduiertenakademie Baden-W\"urttemberg.

\bibliographystyle{mnras}
\bibliography{SpcTidal.bbl}

\label{lastpage}

\end{document}